# Concurrent Magnetic and Metal-Insulator Transitions in Eu$_{1-x}$Sm$_x$B$_6$ Single Crystals

**S. Yeo**

*Korea Atomic Energy Research Institute, 150 Dukjin-Dong, Yuseong-Gu, Daejeon, Republic of Korea*

**J. E. Bunder and Hsiu-Hau Lin**

*Department of Physics, National Tsing-Hua University, Hsinchu 300, Taiwan*

*National Center for Theoretical Sciences, Hsinchu 300, Taiwan*

**Myung-Hwa Jung\* and Sung-Ik Lee**

*Department of Physics, Sogang University, Seoul 121-742, Republic of Korea*

The effects of magnetic doping on a EuB$_6$ single crystal were investigated based on magnetic and transport measurements. A modest 5% Sm substitution for Eu changes the magnetic and transport properties dramatically and gives rise to concurrent *antiferromagnetic* and metal-insulator transitions (MIT) from *ferromagnetic* MIT for EuB$_6$. Magnetic doping simultaneously changes the itinerant carrier density and the magnetic interactions. We discuss the origin of the concurrent magnetic MIT in Eu$_{1-x}$Sm$_x$B$_6$.

\*corresponding author: mhjung@sogang.ac.kr

Electrons, the building blocks for condensed matter physics, have two fundamental physical quantities: charge and spin. The subtle interplay between these two quantities is not only useful for applications such as spintronics, but also presents an exciting challenge to understand how they are intertwined. One enchanting example is the search for magnetic polarons (MPs) [1, 2] where charge carriers are accompanied by a local magnetic polarization and possibly distortions of a nearby crystal lattice. The percolation of MPs leads to concurrent ferromagnetic transition and MIT, showing that the magnetic and the transport properties are intrinsically entangled. In fact, MPs play an important role in the low-density region, where spatial fluctuations overwhelm thermal fluctuations. This interesting phenomenon has been proposed/observed in various contemporary condensed matter systems, such as high-Tc superconductors[3], colossal magnetoresistance materials[4-6], and diluted magnetic semiconductors[7]. Though various experiments indicate the existence of MPs[4, 5, 8], it is still unclear how MPs evolve as the density of electrons ($n_e$) increases[9]. Thus, a systematic study for doping effects in a MP system is an interesting and challenging task.

Typical MP behavior has been observed in Eu-based compounds and in some perovskite manganites. Among the former compounds, $EuB_6$ is one of the best candidates because the intrinsic $n_e$ is low[10] and because the valence states of $Eu^{2+}$ produce local magnetic moments. In addition, unlike in perovskite manganites, in $EuB_6$ lattice distortions, such as the Jahn-Teller effect, are small and do not affect the magnetic and transport properties. Concurrence of the ferromagnetic transition and MIT is sensitive to $n_e$ and detailed studies on a non-magnetic doping such as $Eu_{1-x}Ca_xB_6$ and $Eu_{1-x}La_xB_6$ have been performed[11, 12]. Increasing $n_e$ suppresses the ferromagnetic order and antiferromagnetism appears at about 30 % doping. However, the underlying mechanism remains unclear because $n_e$ is still too low for the carrier-mediated RKKY interaction to turn antiferromagnetic. Inspired by this intriguing puzzle, we take a slightly different route with $Eu_{1-x}Sm_xB_6$ single crystals.

In contrast to previous investigations, Sm not only serves as an efficient dopant, changing the localized electrons into itinerant carriers, but also enhances the local antiferromagnetic coupling since $Sm^{3+}$ with S = 5/2 is a magnetic dopant. This enhancement is easily understood from the fact that $SmB_6$ is a Kondo insulator with a strong antiferromagnetic coupling. These two factors dramatically drive the original MP percolation transition into another category: a concurrent *antiferromagnetic transition* with MIT. It is rather surprising that, by just 5% magnetic doping, the system can be tuned to exhibit MIT driven by different magnetic orderings. In the remaining parts of the Letter, we would like to address how the two types of concurrent magnetic MIT's arise from both experimental and theoretical aspects.

The single crystals used for the study were grown by using the Al flux method. The powder x-ray diffraction pattern obtained by using a Rigaku x-ray diffractometer shows a single-phase cubic $CaB_6$ structure (space group: $Pm3m$) for all $Eu_{1-x}Sm_xB_6$ crystals. In this structure, alkali-earth or rare-earth elements occupy the corner of a cubic structure and six borons make up the octahedron at the center of the cubic. The magnetic properties of the crystals were examined along the cubic axis in a commercial SQUID magnetometer. The resistivity ($\rho$) was measured with a standard four-probe technique and the Hall coefficients were obtained with a Quantum Design PPMS at temperatures between 2 and 300 K. In order to remove the longitudinal magnetoresistance, we also performed Hall measurements in opposite fields.

The magnetization and $\rho$ curves for $EuB_6$ are shown in Fig. 1(a). It clearly displays a ferromagnetic transition at around 15 K. Furthermore, the effective magnetic moment obtained by fitting the Currie-Weiss law at high temperatures for the undoped sample is about 8.1 $\mu_B$, which agrees with the moment carried by $Eu^{2+}$. The $\rho$ data shows an accompanying drop near the ferromagnetic phase transition. The concurrent transitions can be understood within an MP percolation scenario where the percolated magnetic cluster provides the major conducting channel. Since the MP scenario often works when the itinerant carrier density is low or the disordered potentials are strong, it is interesting to explore how the concurrent phase transitions are modified when more itinerant carriers are injected into the sample.

It is rather remarkable that the ferromagnetism is greatly suppressed with a humble amount of Sm doping. As shown in Fig. 1(b), at just $x = 0.05$, the typical antiferromagnetic behavior is already transparent. The Neel temperature ($T_N$) can be determined from the cusp in the magnetization which approaches 13 K for $x = 0.5$, as demonstrated in the inset of the figure. Note that no significant deviations were observed in the field-cooled and zero-field-cooled measurements (not shown here), suggestive of the absence of spin glass states. It is rather interesting that $\rho$ also shows a MIT at ~ $T_N$. Note that the relatively flat $\rho$ curve at high temperatures indicates the spin-flip scattering is weak. Below $T \sim 40$ K, $\rho$ increases on cooling, suggestive of an insulating behavior. However, below $T_N$, $\rho$ decreases on cooling, indicating a metallic behavior. Moreover, as soon as the antiferromagnetic signature is washed out by the external magnetic field, the sign change for the temperature derivatives of $\rho$ also disappears.

Fig. 2 shows the M-H curves for $0 \leq x \leq 0.5$. For EuB$_6$, the ferromagnetic signature is clear. The Sm doping causes a decrease of the M-H curve slope because the ferromagnetism is suppressed by the growing antiferromagnetic order. For instance, at $x > 0.2$ doping levels, the M-H curves are completely linear showing no sign of ferromagnetic correlations. However, it is worth mentioning that at intermediate doping, such as $x = 0.05$, the saturation of the magnetic moment is still visible at strong fields, indicating that the antiferromagnetic order is not very robust.

The Hall resistivity ($\rho_H$) for $0 \leq x \leq 0.05$ in Fig. 3 shows that the electron density increases dramatically with $x$ and the negative slopes show that the charge carriers are electrons. Moreover, the linear $H$ dependence suggests tiny anomalous Hall coefficients due to small spin-orbit couplings, as is observed in other Eu compounds[13]. For $x = 0.05$, the density extracted from the Hall measurement is about $6 \times 10^{20}$ cm$^{-3}$. It is insightful to convert the density into $x_{\text{itin}}$, the average number of *itinerant* carriers per unit cell. Since the Eu spins form a simple cubic lattice with a = 4.184 Å, one obtains $x_{\text{itin}} \sim 0.043$, which is quite close to $x = 0.05$. This means that Sm is a very efficient charge dopant with 86% of the donated electrons entering the band and becoming itinerant. Note that $n_e$ is almost independent of the temperature though $\rho$ shows MIT, suggesting that the concurrent MIT in Eu$_{1-x}$Sm$_x$B$_6$ can be explained by a percolation scenario.

In order to verify this scenario, we estimate the effective exchange coupling using the self-consistent Green's function method that includes both the kinematics and spatial fluctuations of the spin waves appropriately[14, 15]. The proposed model Hamiltonian contains three key ingredients: $H = H_s + H_c + H_{cs}$. $H_s$ is a direct exchange coupling between localized spins, written by $H_s = \sum_{n,m} J_{nm} \vec{S}_n \cdot \vec{S}_m$ where $J_{nm}$ denotes the direct exchange interaction between localized spins at sites $n, m$ in a simple cubic lattice. $H_c$, and $H_{cs}$, written by $H_c = \int \frac{d^3k}{(2\pi)^3} \varepsilon(k) c_\alpha^*(k) c_\alpha(k) + \int d^3r V_{imp}(r) c_\alpha^*(r) c_\alpha(r)$ and $H_{cs} = \int d^3r J(r) \vec{S}(r) \cdot \vec{s}(r)$, stand for the kinetic and the (disordered) potential energies of the charge carriers, and the interaction between the spin density of the charge carriers and the localized spin, respectively. $c_\alpha^*(r)$, and $c_\alpha(r)$ are the creation and the annihilation operators for electrons with spin $\alpha$. $\vec{S}(r) = \sum_n \vec{S}_n \delta^3(r - r_n)$ is the localized spin density, $\vec{s}(r) = \frac{1}{2} c_\alpha^*(r) \sigma_{\alpha\beta} c_\beta(r)$ is the spin density of the charge carriers, and $J(r)$ is the exchange interaction between these two kinds of spin densities. For the system we consider here, the localized spin density is $n_S = 13.65$ nm$^{-3}$ (calculated using the lattice constant of the cubic lattice, a=4.184 Å). Typical values of the exchange coupling between the localized and itinerant spin densities, $J = 20$-$40$ meV nm$^3$, is chosen for the numerical calculations. Unlike the expectation from Weiss mean-field theory, the increased carrier density causes quantum fluctuations and the ferromagnetic coupling mediated by itinerant electrons is greatly suppressed beyond $n_e/n_S = 0.03$. Assuming the 86% efficiency of the Sm doping, it corresponds to $x \sim 0.035$ where the ferromagnetism is expected to disappear. The theoretical estimate is quite consistent with our experimental observations that the magnetic and transport properties are drastically different with 5% magnetic doping.

From previous experiments on non-magnetic doping of EuB$_6$, a weak antiferromagnetic phase emerges when the doping level exceeds 30 %. Since the RKKY interaction cannot turn antiferromagnetic in this density regime, it is therefore reasonable to assume that a weak antiferromagnetic coupling exists between $Eu^{2+}$ spins even in $EuB_6$. In other words, $J_{nm}$ is positive (antiferromagnetic) everywhere and larger in the vicinity of $Sm^{3+}$ spins. Then, the percolation of MP's can explain the appearance of the ferromagnetism in $EuB_6$ where $n_e$ is low. The carrier-mediated exchange coupling is mainly through the formation of MP's. This effective ferromagnetic coupling is a little bit larger than the direct antiferromagnetic coupling so that weak ferromagnetism appears. In this low $n_e$ regime, the magnetic and transport properties are thus well described by the conventional theory for MP percolation. On the other hand, upon Sm doping, $n_e$ increases dramatically, rendering the impurity potentials to be irrelevant after coarse-graining and the Sm dopants tend to enhance a local antiferromagnetic coupling because $SmB_6$ is a Kondo insulator with strong antiferromagnetic correlations and has a slightly shorter bond length. Since the ferromagnetic coupling from itinerant carriers is suppressed, antiferromagnetic clusters start to nucleate in the vicinity of magnetic dopants. Eventually, these clusters grow and percolate, leading to a different type of concurrent magnetic and MIT. In loose terms, the carrier-mediated ferromagnetic interaction from the Hamiltonians ($H_c + H_{cs}$) dominates in $EuB_6$ while the direct antiferromagnetic exchange ($H_s$) takes the lead in $Eu_{1-x}Sm_xB_6$ as long as $x$ is greater than 5 %.

In conclusion, we investigated the magnetic doping effects in $EuB_6$ single crystals by substituting Sm for Eu, which injects electrons into the crystal and modifies the exchange interactions. From both experimental and theoretical perspectives, we have observed interesting concurrent MIT in $Eu_{1-x}Sm_xB_6$ upon magnetic doping. We would like to emphasize that the coupling between the charge carriers and the localized spins makes the system qualitatively different from conventional Heisenberg-like ferro-(antiferro-)magnets. Our measurements clearly demonstrate a concurrent MIT driven by a ferromagnetic phase transition changes to a different MIT driven by the antiferromagnetic one upon magnetic doping.

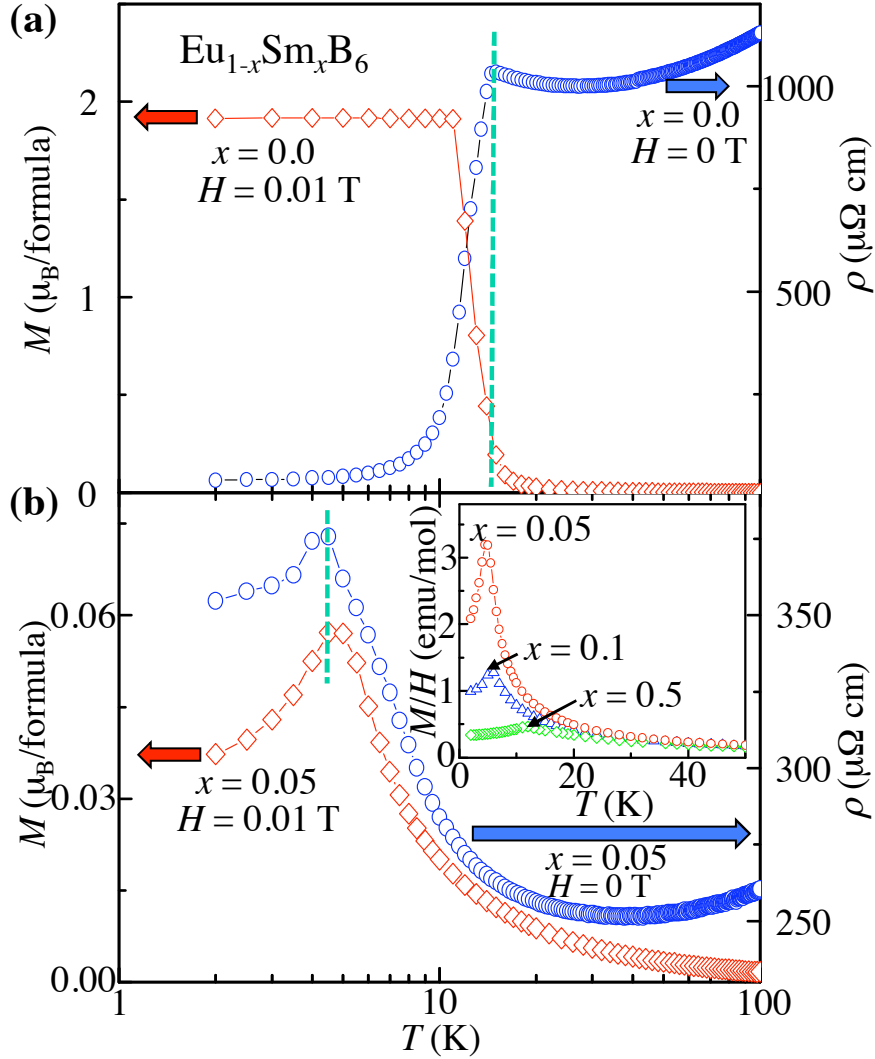

FIG. 1. The temperature dependence of magnetization (at $H = 0.01$T) and $\rho$ (at $H = 0$T) for **(a)** $x = 0$ undoped $EuB_6$ sample and **(b)** $x = 0.05$ magnetically doped sample. Without doping, MIT occurs at the Curie temperature due to the percolation of MP clusters. Upon doping $x = 0.05$, the ferromagnetic transition is replaced by the antiferromagnetic one. However, MIT persists and coincides with the antiferromagnetic transition. The dashed lines are guides to the eyes. The inset displays the magnetization curves for $0.05 \leq x \leq 0.5$ with a clear signature of the antiferromagnetic transition.

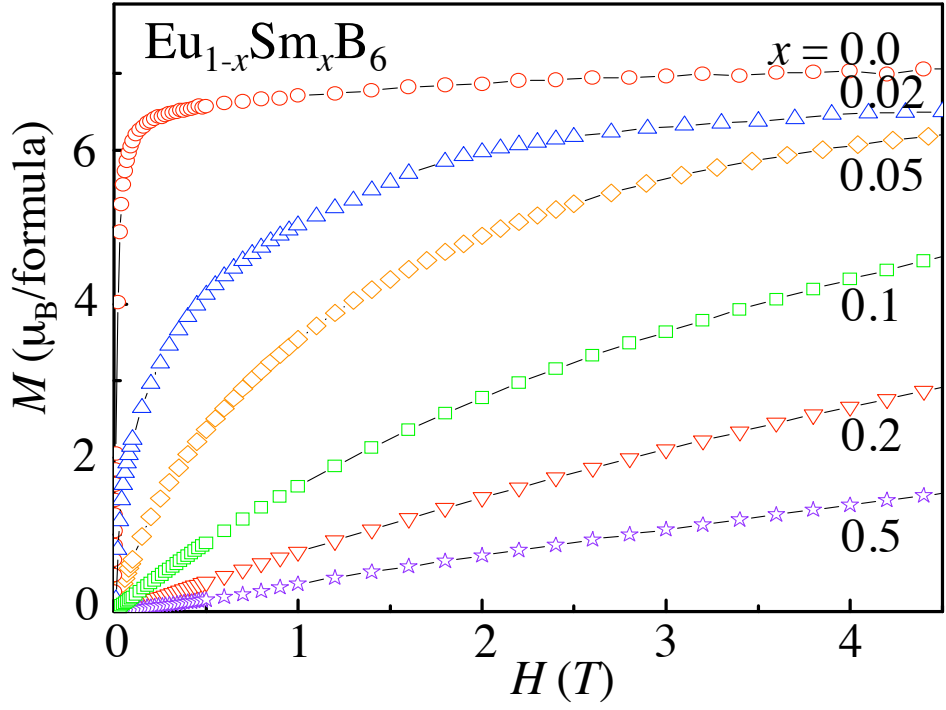

FIG. 2. The magnetization versus the magnetic field at different doping levels for $Eu_{1-x}Sm_xB_6$ obtained at 5 K.

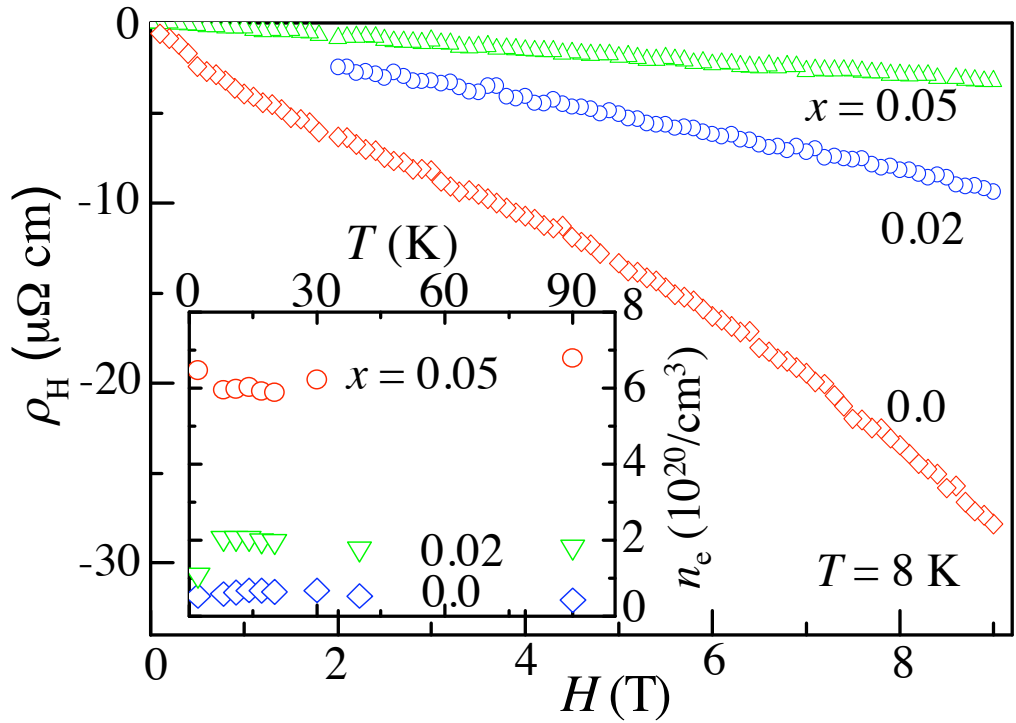

FIG. 3. $\rho_H$ for $x$ = 0.0, 0.02, and 0.05 at 8 K. The negative slope indicates that the itinerant charge carriers are electrons. The inset shows the itinerant carrier density extrapolated from $\rho_H$.